\documentclass[pra,amsmath,amssymb,twocolumn,superscriptaddress,hidepacs]{revtex4}

\usepackage{amsmath,amssymb}
\usepackage[usenames]{color}
\usepackage{amssymb}
\usepackage{grffile}
\usepackage[pdftex]{graphicx}
\usepackage{amsmath, amstext, amssymb, amsfonts, amsxtra}
\usepackage{textcomp}
\usepackage{xspace}
\usepackage{epstopdf}
\DeclareGraphicsExtensions{.eps}
\usepackage[percent]{overpic}
\usepackage{enumerate}
\usepackage{mathbbol}

\newcommand{\e}{\textrm{e}}
\newcommand{\ii}{\textrm{i}}

\renewcommand{\Re}{\mathcal{R}\textrm{\small e}}
\renewcommand{\Im}{\mathcal{I}\textrm{\small m}}

\newcommand{\be}{\begin{equation}}
\newcommand{\ee}{\end{equation}}
\newcommand{\bea}{\begin{eqnarray}}
\newcommand{\eea}{\end{eqnarray}}
\newcommand{\la}{\langle}
\newcommand{\ra}{\rangle}

\begin{document}	

\title{Ground-state properties of the one-dimensional unconstrained pseudo-anyon Hubbard model}
\author{Wanzhou Zhang}
\affiliation{College of Physics and Optoelectronics, Taiyuan University of Technology Shanxi 030024, China
}
\affiliation{C.N. Yang Institute for Theoretical Physics,
State University of New York at Stony Brook, Stony Brook, NY 11794-3840, USA}
\author{Sebastian Greschner}
\affiliation{Institut f\"ur Theoretische Physik, Leibniz Universit\"at Hannover, Appelstr. 2, DE-30167 Hannover, Germany}
\author{Ernv Fan}
\affiliation{College of Physics and Optoelectronics, Taiyuan University of Technology Shanxi 030024, China
}
\author{Tony C Scott}
\affiliation{College of Physics and Optoelectronics, Taiyuan University of Technology Shanxi 030024, China
}
\affiliation{Near Pte Ltd, 35th Floor, UOB Plaza 1, 80 Raffles Place, Singapore 048624}

\author{Yunbo Zhang}
\affiliation{Institute of Theoretical Physics, Shanxi University, Taiyuan 030006, China }
\date{\today}
\begin{abstract}
We study the (pseudo-) anyon Hubbard model on a one-dimensional lattice without the presence of a three-body hardcore constraint.
In particular, for the pseudo-fermion limit of a large statistical angle $\theta\approx\pi$, we observe a wealth of exotic properties including {a first order transition} between different superfluid phases and a {two-component} partially paired phase for large fillings without need of an additional three-body hardcore constraint.
In this limit, we analyze the effect of an induced hardcore constraint, which leads to the stabilization of superfluid {ground states} for vanishing or even small attractive on-site interactions.
For finite statistical angles, we study the unconventional broken-symmetry superfluid peaked at a finite momentum, resulting in an interesting beat phenomenon of single particle correlation functions.
We show how some features of various ground state phases, including an analog of the partially paired phase in the pseudo-fermion limit, may be reproduced in a naive mean field frame.
\end{abstract}
\maketitle

\section{Introduction}

Bosons and fermions are the two types of well-known elementary particles.
By exchanging two bosons (fermions), the wave function  is  symmetric (anti-symmetric), or updated
with a phase factor $e^{i\theta}$, where $\theta=0$ for bosons, and $\theta=\pi$ for fermions.
In low dimensions, particles with other types of quantum statistics, anyons, are possible. Anyons are governed by statistics which are intermediate between those of bosons and fermions. The exchange of two identical {anyons will acquire} a  phase angle
$\theta$, which can be of any value.
Since the 1980s\cite{first}, anyons have attracted much physical interest and
have become a very important concept in condensed matter physics
{including the} fractional quantum Hall effect~\cite{Laughlin1983,Halperin1984,Haldane1991, Camino2005,Kim2005} and topological quantum computing~\cite{Nayak2008,kitaev}.

Experimentally, several schemes have been proposed to search for the anyons in spin or boson models~\cite{kitaev, longguilu,jianweipan, jingzhang, nmr}, or
in cold atoms~\cite{ex5,ex6,ex7,ex8,ex9}.
During recent years, ideas for the realization of (pseudo-) anyons in one-dimensional optical lattices as initially proposed by Keilmann et al.~\cite{Keilmann2011} have attracted considerable interest. Here, a Raman-assisted hopping scheme would allow for the manipulation and engineering of the anyonic exchange statistics in an optical lattice experiment.
Recently, this experimental scheme for the realization of such anyon-Hubbard models (AHM) has been refined~\cite{Greschner2015}, drastically simplified~\cite{Straeter2016} and extended to two-component anyons~\cite{Cardarelli2016}. Typically these models are only valid in the low density regime or impose a three body hardcore constraint, restricting the local particle number per site to $n_{max}=2$. However, there have been proposals for the realization of AHM-like models~\cite{Greschner2014} by means of modulated interactions without this restriction.

\begin{figure}[b]
\includegraphics[width=\columnwidth]{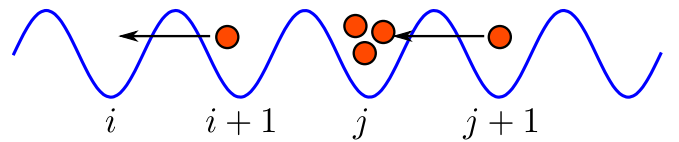}
\caption{The illustration of the conditional effect of $e^{\theta n_i}$,
$n_i=0$,  $e^{\theta n_i}=1$; $n_j=3$, $e^{\theta n_j}\ne1$.}
\label{fig:con}
\end{figure}

While an experimental realization of one-dimensional anyons on a lattice is still lacking, theoretically the physics of AHM has been studied extensively during recent year.
Properties of the hardcore AHM (i.e. with a constraint of local Hilbert space to $0$ or $1$ particles) or low density continuum models, their asymmetric momentum distributions~\cite{Calabrese2007,Patu2007,Hao2008,Hao2009,Calabrese2009}, intriguing particle dynamics~\cite{DelCampo2008,Hao2012} and entanglement properties~\cite{Santachiara2007,Guo2009, Marmorini2016} have all been studied.

Pseudo-AHMs~\cite{Keilmann2011} (or {particular two-component AHMs}~\cite{multicomponent, Cardarelli2016}), where the particles off site obey anyon-like commutation relations and on site act like bosons, have been shown to exhibit a rich phase diagram dependent on the statistical angle $\theta$. While previous studies of
ground-state properties focused on the analysis of the Mott insulator~(MI) to superfluid~(SF) transitions, including  statistically induced MI to SF quantum phase transitions~\cite{Keilmann2011,Forero2016}, properties of momentum distribution~\cite{Tang2015} and expansion dynamics~\cite{Wang2014}, recently in Ref.~\cite{Greschner2015} a wealth of further ground-state phases of
the AHM was found: dimerized phases and an unconventional two-component partially paired (PP) phase were realized for statistical angles $\theta\to \pi$. Since the latter work focused on the AHM in the presence of an artificial {three-body hardcore constraints} due to the proposed experimental realization, it remained unclear {whether or not} the interesting {ground-state} phases, and in particular, the PP phase can arise for the pseudo-AHM without further constraint on the local particle number.

In this work, we fill this gap and study the {ground-state} phase diagram of {the} unconstrained AHM
particularly focusing on statistical angles close to the pseudo-fermion limit $\theta\to\pi$. We discuss three fundamental properties of the AHM: effective statistically induced repulsive interactions, a density-dependent (drift of the) momentum distribution and the emergence of the exotic two-component PP phase.

The starting point is the AHM,
\be
H^\alpha =-t\sum_{i=1}^{L}(\alpha^{\dagger}_{i}\alpha_{i+1}+h.c.)+\sum_{i}h_i\\
\label{ha}
\ee
where $\alpha_i^{\dag} (\alpha_i)$ is the anyon creation (annihilation) operator at site $i$,
$t$ is the single-anyon  hopping amplitude,
$L$ is the lattice size,
and $n_i=\alpha_i^{\dag}\alpha_i$ is the number operator of the anyons on site $i$.
$\alpha_j$ and $\alpha_j^\dagger$ satisfy anyonic commutation relations, $\alpha_j \alpha_k^\dagger - {\rm e}^{-\ii\theta\, {\rm sgn}(j-k)} \alpha_k^\dagger \alpha_j = \delta_{jk}$ and $\alpha_j \alpha_k - {\rm e}^{-\ii\theta\, {\rm sgn}(j-k)} \alpha_k \alpha_j = 0$. It is important to note, that the particles on-site behave like bosons. This means, for example, that even in the pseudo-fermion limit $\theta\to\pi$, more than one particle is allowed on the same lattice site.

In the term $h_i=\frac{U}{2} n_i    (n_i-1)-\mu n_i$, $U$ is the
on-site two-body interaction and $\mu$ is the chemical potential term.
By a Jordan-Wigner transformation~\cite{Keilmann2011},
\be
\alpha_j=b_{j}{\rm e}^{-\ii\theta\sum_{i=1}^{j-1}n_i},
\ee
where $b_j$ is a boson annihilation operator,
the anyon Hamiltonian $H^\alpha$ can be re-expressed as a Bose-Hubbard model with a density dependent phase factor\cite{Keilmann2011}:
\be
H^b =-t\sum_{i=1}^{L}(b^{\dagger}_{i}b_{i+1}e^{i\theta n_i}+h.c.) +\sum_{i}h_i.
\label{BH}
\ee

Fig.~\ref{fig:con} shows the conditional effects of the  density-dependent phase factor caused by $b^{\dagger}_{i}b_{i+1}e^{i\theta n_i}$.
If there are no particles in the site $i$, namely $n_i=0$, then the phase factor is still given by $e^{i\theta n_i}=1$.
The situation becomes different for a soft-core Bose-Hubbard model, allowing of more than one particle on each site $n_{max}>1$.
If three particles already exist in the site $j$ as shown in the example in Fig.~\ref{fig:con}, the phase factor becomes
$e^{i\theta n_j}=e^{i3\theta}$.

The paper is {structured} as follows: In section~\ref{sec:PP}, we present strong indications of an emergent two-component PP phase at large fillings $\rho\gtrsim 1.5$  by means of density matrix renormalization group~(DMRG) simulations~\cite{dmrg1}. This phase can be understood as the presence of both an atomic and a paired superfluid component. We also show how some feature of the models may be understood within an intuitive mean field~(MF) and dilute limit picture.
In section~\ref{sec:eff_int}, we show the emergence of an effective Pauli exclusion principle induced by the anyonic exchange statistics which leads to a stabilization of particle density for vanishing interactions. Finally, in section~\ref{sec:bsf}, we analyze the asymmetric momentum distribution. We study the crossover and transitions between superfluid phases condensed at a momentum $0<\theta<\pi$ with broken reflection symmetry (SF$_Q$ or broken symmetry superfluid, BSF) and analyze the single particle correlation function. Concluding comments are made in Sec.~\ref{sec:con}.

\section{The PP phase for pseudo-fermions} \label{sec:PP}

For the constrained AHM, the PP phase was described in Ref.~\cite{Greschner2015} and studied extensively using DMRG simulations. It may be extended to quasi-1D ladder models~\cite{Mishra2016} and also {a} variant of the hardcore {two-component AHM} may exhibit a similar multicomponent PP phase~\cite{Cardarelli2016}. Here we present detailed numerical evidence for the emergence of this phase for the unconstrained model~\eqref{ha} in the limit $\theta\to \pi$.

\subsection{DMRG results}

In the following, we study the one-dimensional~(1D) AHM~\eqref{ha} by
numerically exact DMRG calculations, which are performed with open and periodic boundary conditions keeping up to $L=160$ sites and $m=1000$ matrix-states~\cite{dmrg1}. We make sure that our results are independent of the system size and the cut-off of the local bosonic Hilbert-space $n_{max}$.

\begin{figure}[tb]
\centering
\includegraphics[width=\linewidth]{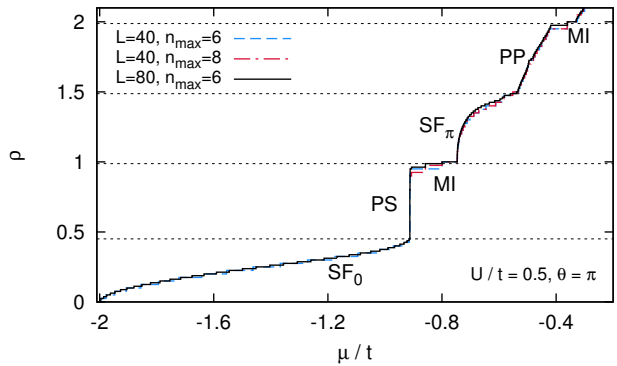}
\caption{Equation of state $\rho=\rho(\mu)$ for the AHM for
{a} statistical angle $\theta=\pi$ ($U=0.5 t$, DMRG). Several different phases and phase transitions may be observed. Around $\rho\approx 1.5$, we observe a marked kink in the $\mu$-$\rho$ curve signaling a change in the number of gapless excitations and we identify the region for $1.5 \lesssim \rho \lesssim 2$ with the PP phase.}
\label{fig:dmrg_pi_mag}
\end{figure}

\begin{figure}[tb]
\centering
\includegraphics[width=\linewidth]{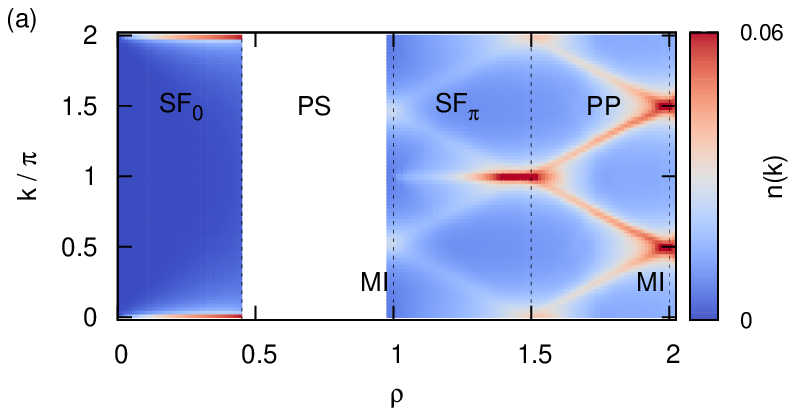}
\includegraphics[width=\linewidth]{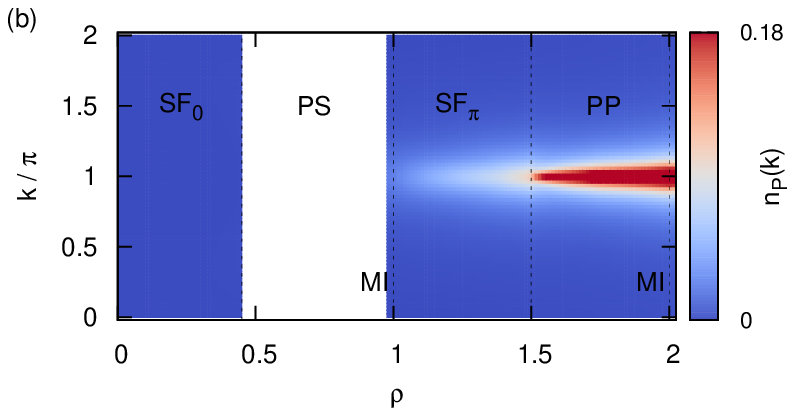}
\caption{(Bosonic) momentum distribution $n(k)$ (a) and pair momentum distribution $n_P(k)$ (b) for the AHM (DMRG, $L=80$, $n_{max}=6$) as a function of the density for {a} statistical angle $\theta=\pi$. We do not display data for the
the phase separation (PS) region.}
\label{fig:dmrg_pi_mom}
\end{figure}

\begin{figure}[tb]
\centering
\includegraphics[width=\linewidth]{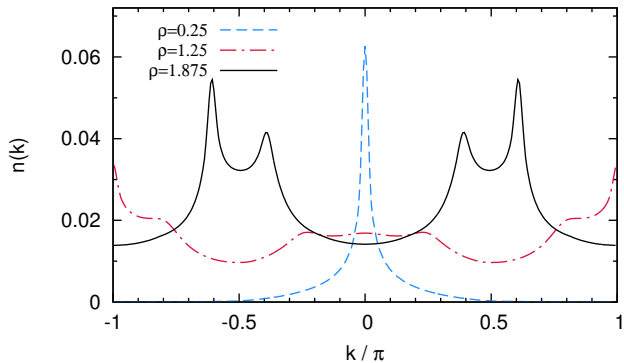}
\caption{(Bosonic) momentum distribution $n(k)$ for the AHM (DMRG, $L=80$, $n_{max}=6$, $U=0.5t$) for {a} statistical angle $\theta=\pi$ and densities $\rho=0.25$ (SF$_0$ phase), $\rho=1.25$ (SF$_\pi$ phase) and $\rho=1.875$ (PP phase).}
\label{fig:dmrg_pi_mom_cut}
\end{figure}

In Fig.~\ref{fig:dmrg_pi_mag}, we show the equation of state $\rho=\rho(\mu)$ for $\theta=\pi$ and small repulsive interactions $U=t/2$.
We may identify several gapless and gapped phases. Mott-insulating phases
with a {mass gap} and a vanishing compressibility are characterized by a horizontal plateau in the $\mu-\rho$ curve.  As discussed in Ref.~\cite{Keilmann2011}, due to the statistical angle $\theta\to\pi$, we may still observe MI phases for this small interaction strengths  at $\rho=1$ and $\rho=2$. Note that the step-like behavior of the plateaus is an effect of the finite system size and open boundary conditions.

For incommensurate particle fillings $\rho$, we observe several gapless quantum phases. For $0<\rho\lesssim 0.5$ and $1<\rho\lesssim 1.5$ ordinary (one component) Luttinger-liquid phases are stabilized. As their momentum-distribution function (see Fig.~\ref{fig:dmrg_pi_mom})
\be
n(k)=\frac{1}{L}\sum_{i,j}\langle b_i^{\dagger}b_{j}\rangle e^{ik(i-j)},
\label{nk}
\ee
is peaked at $k=0$ and $k=\pi$ respectively, we call these quasi-superfluid phases SF$_0$ and SF$_\pi$.

For larger densities $0.5\lesssim \rho\lesssim 1$, the DMRG analysis shows that interplay between interactions and exchange statistics leads to an instability of the system, characterized by a large macroscopic jump in density and a separation of phases (PS).

Around $\rho\sim 1.5$, we observe a pronounced kink in the $\mu$-$\rho$ curve of Fig.~\ref{fig:dmrg_pi_mag} which indicates a commensurate-incommensurate transition in which the number of gapless modes changes, corresponding to the transition from the one-component SF$_{\pi}$ to the two-component partially paired (PP) phase as discussed in Ref.~\cite{Greschner2015}. This PP phase can be understood as a phase of both a gas of atomic pseudo-fermions and a quasi-condensate of pairs. In the appendix~\ref{sec:app2particle}, we discuss a simplified model for this characterization.

The most characteristic signature of the PP phase is given by its (bosonic)
momentum distribution function, which exhibits a characteristic multi-peak structure as shown in Figs.~\ref{fig:dmrg_pi_mom}~(a) and \ref{fig:dmrg_pi_mom_cut}. The largest peak is located at incommensurate values $0<k<\pi$. Interestingly, the SF$_\pi$ phase also exhibits several local maxima, in addition to the distinct peak at $k=\pi$. We also evaluate the pair-momentum distribution
\begin{align}
n_p(k) = \frac{1}{L} \sum_{i,j} \e^{\ii(i-j)k} \langle \left(b_i^\dagger\right)^2 b_j^2\rangle \;,
\end{align}
which is shown in Fig.~\ref{fig:dmrg_pi_mom}~(b). As conjectured in the dilute limit analysis (see appendix~\ref{sec:app2particle}), we see the formation of a sharp peak at $k=\pi$, indicating a quasi-condensation of pairs in the PP phase.

\begin{figure}[tb]
\centering
\includegraphics[width=\linewidth]{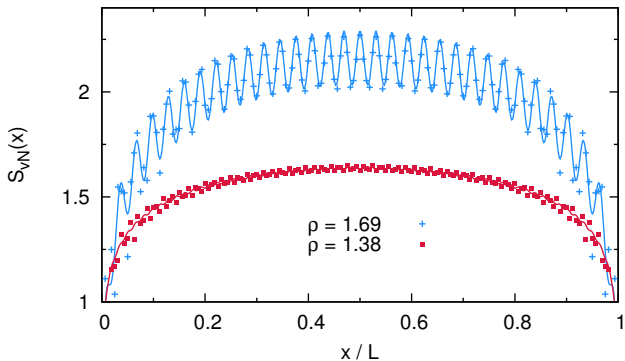}
\caption{Examples of the scaling of the von Neumann block entanglement entropy $S_{vN}$ as a function of the bipartition position of the system for two different densities $\rho=270/160$ (upper curve, PP phase) and $\rho=220/160$ (lower curve, SF$_\pi$ phase) ($L=160$ sites, $U=0.5 t$, $\theta=\pi$). The DMRG data points are fitted by the Calabrese-Cardy formula Eq.~\eqref{eq:CC} assuming additional higher order oscillatory terms and a central charge $c=1$ for $\rho=220/160$ and $c=1+1=2$ (using the convection of e.g. Ref.~\cite{Hikihara2010}) for $\rho=270/160$.}
\label{fig:dmrg_pi_c}
\end{figure}

In order to further verify the two-component character of the PP phase, we study the scaling of the von Neumann entanglement entropy $S_{\mathrm{vN}}= -\mathrm{tr}\left( \rho_x \ln\rho_x \right)$, where $\rho_x$ is the reduced density matrix of a subsystem of length $x$ embedded in a chain of a
finite length $L$. One can relate the scaling of the entanglement entropy to the central charge $c$ of the system, which basically counts the number of gapless excitations, as described by the Calabrese-Cardy formula~\cite{Holzhey1994, kitaev2003, korepin2004, Calabrese2004}
\be
S_{\mathrm{vN}}(x) = \frac{c}{6} \ln\left[ \frac{L}{\pi} \sin\left(\frac{\pi}{L}x\right) \right] + \cdots \,\,,
\label{eq:CC}
\ee
The ellipsis contains non universal constants and higher order oscillatory terms due to the finite system size and open boundary conditions.
In Fig.~\ref{fig:dmrg_pi_c}, we show examples of $S_{vN}(l)$ for different densities corresponding to the SF$_{\pi}$ and the PP region. A fit to Eq.~\eqref{eq:CC}, including an assumed higher order oscillatory part, shows the consistency  with a central-charge $c=1$ in the SF- and $c=1+1=2$ in the PP region (here we follow the convention of e.g. Ref.~\cite{Hikihara2010}, in which a $c=1+1$ phase can be understood as a phase of two critical modes).

\begin{figure}[tb]
\centering
\includegraphics[width=\linewidth]{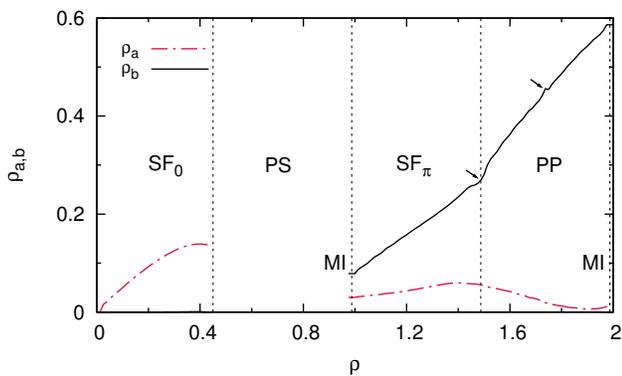}
\caption{Sum of local single- $\rho_a$ and two-particle correlations $\rho_b$ (DMRG, $L=80$, $n_{max}=6$, $U=0.5t$) as defined in the main text. Arrows indicate an interesting plateau-like substructure in the $\rho_b$ curve.}
\label{fig:dmrg_pi_cc}
\end{figure}

In Ref.~\cite{Greschner2015}, the atom and doublon dimerizations,
\begin{align}
\rho_a &= \frac{2}{L} \sum_{L/4<i<3L/4} \left<b_i^\dagger b_{i+1}\right> ,\nonumber\\
\quad \rho_b &= \frac{2}{L} \sum_{L/4<i<3L/4} \left<(b_i^\dagger)^2 (b_{i+1})^2\right> \;,
\end{align}
have been shown to be a good probe for the PP and SF phases. In Fig.~\ref{fig:dmrg_pi_cc}, we present $\rho_a$ and $\rho_b$ as a function of the density for the parameters of Fig.~\ref{fig:dmrg_pi_mag}.
This picture shows that the low density SF$_0$ phase is a LL phase of almost hardcore single particles ($\rho_b\approx 0$). This is not true for the SF$_\pi$ phase for $1\lesssim \rho \lesssim 1.5$. Here, we observe that $\rho_a,\rho_b >0$ and are finite as well as a linear increase of both of these quantities with density.  However, the PP phase is characterized by a enhanced increase of $\rho_b$ while $\rho_a$ decreases.

Apparently, the $\rho_{a(b)}-\rho$ and the $\mu$-$\rho$ curves of Fig.~\ref{fig:dmrg_pi_mag}, exhibit some kind of interesting substructure. One may observe several small plateau-like steps at certain commensurate fillings $\rho\approx 1.5$ and $\rho\approx 1.75$ which could indicate the formation of pair-crystals due to an effective pair-pair interaction. Similar plateaus at fractional fillings have also been observed in Ref.~\cite{Keilmann2011} for a trapping potential. This interesting phenomenon will be studied elsewhere.

In Fig.~\ref{fig:pd_p1.0}~(a), we summarize the DMRG results in a phase diagram for $\theta=\pi$ as a function of the hopping $t/U$ and the chemical potential $\mu/U$. The extent of the PP phase shrinks drastically with increasing interaction strength. Indeed, as shown in the $\mu-\rho$ curve of Fig.~\ref{fig:MF_cut}, the PP phase is hardly visible at a large interaction strength $U=2t$ and the transition to the SF$_\pi$ phase apparently becomes of first order (while, interestingly, the region of phase separation (PS) below $\rho<1$ vanishes and a direct phase transition between the SF$_0$ phase and the MI at $\rho=1$ is found). The properties of the gapless phases for fillings $\rho>2$ are not studied in this work.

\subsection{The PP phase in the mean field approximation}

In the following we will discuss a naive mean field approach to gain intuitive insight into the physics of model~\eqref{BH}.
It is important to note, that the mean field analysis is (a priori)
unreliable in particular in strongly-correlated one-dimensional systems as it e.g. may incorrectly predict a spontaneous breaking of continuous symmetries as forbidden by the Hohenberg-Mermin-Wagner theorem~\cite{Mermin1966, Pitaevskii1991}.
Nevertheless, the mean field ansatz can be used to reproduce certain features of the model on a qualitative level, which may be compared to the exact (DMRG) numerical treatment.

Keilmann et al.~\cite{Keilmann2011} perform a simple mean field approximation to describe the modification of the MI-SF boundary due to the statistical angle. In accordance with the above statement, the true 1D superfluid phase does not exhibit a spontaneously broken $U(1)$-symmetry and long range order, but is a Luttinger-liquid phase with algebraically decaying correlation functions. Hence, such MF statements have to be complemented by exact analytical and numerical methods such as DMRG (see for example Refs.~\cite{Keilmann2011, Forero2016} for a comparison between the mean field and DMRG results for the case of Mott-insulator boundaries of the AHM).
In Ref.~\cite{Tang2015}, a modified Gutzwiller-MF techniques is used for the analysis of the momentum-distribution of the 1D AHM model.

\begin{figure*}[tb]
\centering
\includegraphics[width=0.32\linewidth]{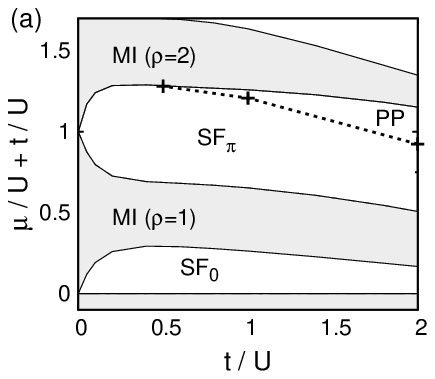}
\includegraphics[width=0.32\linewidth]{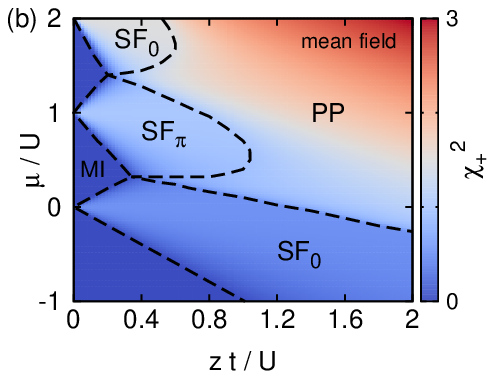}
\includegraphics[width=0.32\linewidth]{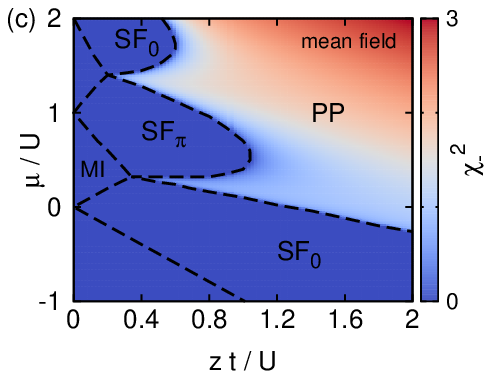}
\caption{(a) Ground-state phase diagram of model~\eqref{ha} for $\theta=\pi$ as obtained by (a) DMRG calculations as a function of $t/U$ and $(\mu-t)/U$ (up to $n_{max}=6$ bosons) and (b)-(c) mean field approach as a function of $zt/U$ and $\mu/U$ ($n_{max}=10$ bosons). Shadings indicate (b) the total superfluid density $\chi_+^2$ and (c) the relative SF density $\chi_-^2$.}
\label{fig:pd_p1.0}
\end{figure*}

In order to put the analysis on a more solid ground, we may consider the following possible generalization of model~\eqref{BH} to a multi-dimensional bipartite lattice with coordination number $z$
\be
H^{z} =-t\sum_{\la ij \ra}(b^{\dagger}_{i} e^{\ii \theta n_i} b_j+h.c.) + \frac{U}{2}\sum_{i} n_i (n_i-1).
\label{BH2d}
\ee
Here, the summation $\la ij \ra$ runs over nearest neighbor sites. Obviously, the intriguing interpretation of model~\eqref{BH2d} as a model for anyons Eq.~\eqref{ha} is only possible in the limit of one physical dimension. However, the correlated hopping Bose-Hubbard model~\eqref{BH2d} in e.g. 2D could be engineered in cold-atom experiments with minimal additional experimental effort as compared to the 1D case (see e.g. Ref.~\cite{Greschner2015DDSM} for a similar discussion) and - as the mean field analysis suggests - could exhibit intriguing many-body physics, some of which can be related to the physics of the AHM in
one dimension.

Anticipating the discussion in section~\ref{sec:bsf} for $0<\theta<\pi$, our mean field ansatz does not predict the SF$_Q$ phases as observed in the DMRG simulation and the ansatz should be extended (compare to e.g. Ref.~\cite{Tang2015}) to capture such features.
However, remarkably, in the limit of $\theta\to\pi$ our mean field results provide a qualitatively accurate picture which also captures non-trivial effects (e.g. a hardcore repulsion of particles as discussed in section~\ref{sec:eff_int}) and quantum phases (MI, SF$_0$, SF$_\pi$ and even PP phase) as we will show in the following. A better quantitative agreement of the phase diagram obtained by the MF approach with the 1D DMRG simulation is found only in the limit of very strong interactions $t\ll U$ \cite{Keilmann2011}.

Following Keilmann et al.~\cite{Keilmann2011}, the density dependent hopping
$ b_j^{\dagger}e^{i\theta n_j}b_{j+1}= c_{j}^{\dagger}b_{j+1}$
is decoupled as
\[ c_{j}^{\dagger}b_{j+1}\approx-\Psi_{2,j}^*\Psi_{1,j+1}+\Psi_{2,j}^{*}b_{j+1}+c_j^{\dagger}\Psi _{1,j+1}~,
\]
where the order parameters are introduced as $\Psi_{1,j}=\la b_j\ra$ and $ \Psi_{2,j}=\la c_j\ra $.

Assuming a homogeneous solution $\Psi_{1}=\la b_j\ra=\la b_{j+1}\ra$, $\Psi_2=\la c_j \ra=\la c_{j+1} \ra $, the decoupled Hamiltonian may be written as
\begin{align}
H=&- z t(\Psi_2b^{\dagger}+\Psi_2^{*}b+\Psi_1c^{\dagger}+\Psi_1^{*}c-\Psi_{1}^*\Psi_{2}-\Psi_2^{*}\Psi_1)  \nonumber\\
& + \frac{U}{2} n(n-1) -\mu n \;.
\label{hs}
\end{align}
This system has to be solved self-consistently for $\Psi_{1}$, $\Psi_{2}$. The solution minimizes the energy functional
$E(\Psi_1, \Psi_2)$,
where
$E(\Psi_1, \Psi_2)$
is the lowest eigenenergy of $H$ for a given set of order parameters $\Psi_1$ and $\Psi_2$~\cite{Wagner2012}.

One may easily extend the mean field ansatz to larger unit cells. E.g. for a two site $L=2$ {unit cell}, we use subscripts $A$ and $B$ to distinguish
the physical quantities $\Psi_{1}$ and $\Psi_{2}$, on the different sublattices, such as $\Psi_{1A}$, $\Psi_{1B}$,  $\Psi_{2A}$, and $\Psi_{2B}$.
We define the average density of atoms on both sublattices as
$\rho_{A}=\langle n_A \rangle$ and $\rho_B =\langle n_B\rangle$.
Combining Eq.~(\ref{hs}) and the definitions of order parameters, we obtain the local Hamiltonian on the sublattice A and thus
\begin{equation}
\begin{aligned}
H_A&=-\frac{zt}{2}[c^{\dagger}_{A}\Psi_{1B}+c_{A}\Psi^{*}_{1B}+b_{A}\Psi^{*}_{2B}+b^{\dagger}_{A}\Psi_{2B}\\
&-\frac{1}{2}(\Psi^{*}_{2A}\Psi_{1B}+\Psi_{2A}\Psi^{*}_{1B}+\Psi^{*}_{2B}\Psi_{1A}+\Psi_{2B}\Psi^{*}_{1A})]\\
&+\frac{U}{2} n_{A}(n_A-1)-\mu n_{A},
\label{ha2}
 \end{aligned}
\end{equation}
and the Hamiltonian on $H_B$ is
\begin{equation}
\begin{aligned}
H_B&=-\frac{zt}{2}[b^{\dagger}_{B}\Psi_{2A}+b_{B}\Psi^{*}_{2A}+c^{\dagger}_{B}\Psi_{1A}+c_{B}\Psi^{*}_{1A}\\
&-\frac{1}{2}(\Psi^{*}_{2A}\Psi_{1B}+\Psi_{2A}\Psi^{*}_{1B}+\Psi^{*}_{2B}\Psi_{1A}+\Psi_{2B}\Psi^{*}_{1A})]\\
&+\frac{U}{2} n_{B}(n_{B}-1)-\mu n_{B}
\label{hb2}
\end{aligned}
\end{equation}
Again Eqs.~(\ref{ha2}) and (\ref{hb2}) have to be solved self-consistently.
A slightly different mean field approach has been employed recently in Ref.~\cite{Tang2015}.

\begin{figure}[tb]
\centering
\includegraphics[width=\columnwidth]{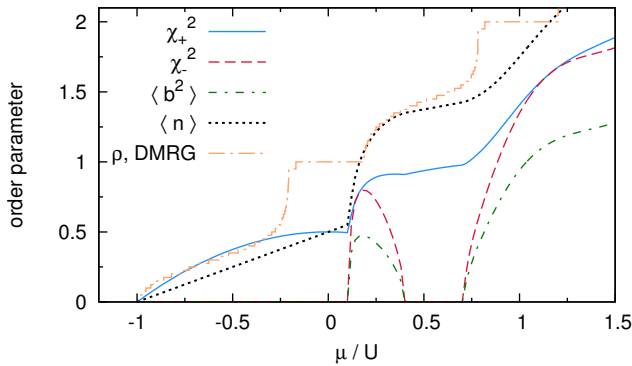}
\caption{Superfluid order parameters, average density and pair-density (obtained by the mean field approximation) for $\theta=\pi$ and $z t= U$ as a function of $\mu/U$. We also show the $\rho=\rho(\mu)$ curve as obtained by DMRG simulations for the same parameters ($U=2t$, $\theta=\pi$) which shows that the
mean field results do not capture correctly the physics of the AHM on a quantitative level.}
\label{fig:MF_cut}
\end{figure}

\begin{figure}[tb]
\centering
\includegraphics[width=0.49\columnwidth]{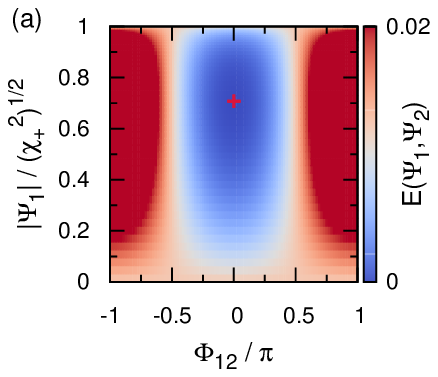}
\includegraphics[width=0.49\columnwidth]{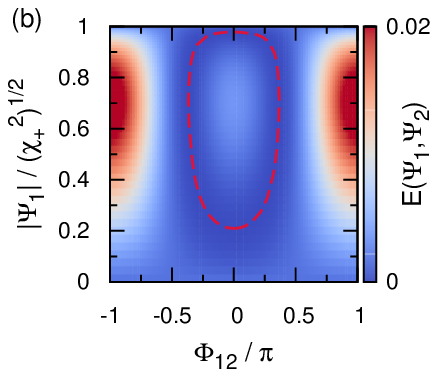}
\caption{{\bf Mean field} energy functional $E(\Psi_1,\Psi_2)$ as {a function} of the relative phase $\Phi_{12}$ and amplitude $|\Psi_1|$ for (a) $\mu/U=0.16$ (SF$_0$ phase) and (b) $\mu/U = 0.19$ (PP phase) for $zt=0.8 U$ (for fixed $|\Psi_1|^2+|\Psi_2|^2 \equiv \chi_+^2$). The red symbol and the dashed line mark the minima as obtained by a self-consistent solution of the mean field equations. We want to stress, that these properties have to be understood as a peculiarity of the mean field solution and have no meaning on the level of the true 1D-AHM~\eqref{ha} studied by DMRG simulations.}
\label{fig:emf_j0.4}
\end{figure}

Within the mean field framework, we mainly find solutions with $|\Psi_{1}|^2=|\Psi_{2}|^2$, the phase between $\Psi_{1}$ and $\Psi_{2}$ is fixed to a particular value $\arg(\Psi_2^{*}\Psi_1) \sim \theta \rho$ and $||\Psi_{1}|^2-|\Psi_{2}|^2| = 0$. For a larger unit cell $L=2$, the relative phase between different sites is found to be $\arg(\Psi_{1A}^{*}\Psi_{1B}) = 0$ or $\pi$, corresponding to {the} SF$_0$ and SF$_\pi$ phases.

A self-consistent {ground-state} solution with $\Psi_1 = \Psi_2\equiv \Psi$ can be found analytically for $\theta\to \pi$. Then, due to the {assumption of a} density dependent correlated hopping $b^\dagger (-1)^n b$, the Hamiltonian reduces to a block-diagonal form with the hopping term only coupling
Fock states $|0\ra$ and $|1\ra$, $|2\ra$ and $|3\ra$, etc. So for the case of $n$ and $n+1$ particles, the Hamiltonian is given by
\begin{align}
&H_{n,n+1} = \nonumber\\
&\left(
\begin{array}{cc}
z |\Psi|^2 t-\mu n+\frac{(n-1) n U}{2} & -z \Psi t \sqrt{1+n} \\
-z \Psi^* t \sqrt{1+n} & z |\Psi|^2 t-\mu (1+n)+\frac{n (1+n) U }{2} \\
\end{array}
\right)
\end{align}
which gives rise to the solution
\begin{align}
&\Psi_{n,n+1} \to \frac{\sqrt{(zt(1+n))^2 -\mu^2+2 \mu n U-n^2 U^2}}{2 zt \sqrt{1+ n}}
\label{eq:MF_hardcore_psi}
\end{align}

Interestingly, we find a second type of {ground-state} {solution, namely} the relative phase and amplitude of $\Psi_{1}$ and $\Psi_{2}$ may fluctuate, while the total amplitude,
\begin{align}
\chi_{+}^2 \equiv |\Psi_{1}|^2+|\Psi_{2}|^2 ,
\end{align}
is fixed. In Fig.~\ref{fig:emf_j0.4}, we illustrate the energy functional $E(\Psi_1,\Psi_2)$ Eq.~\eqref{hs} as a function of the amplitude $|\Psi_1|$ and the relative phase $\Phi_{12} = {\rm arg}(\Psi_1)-{\rm arg}(\Psi_2)$. Its minima correspond to the set of self-consistent solutions of the mean field equations~\eqref{hs}. For the SF phase, there is only one minimum (see Fig.~\ref{fig:emf_j0.4}~(a)) corresponding to the case $|\Psi_{1}|=|\Psi_{2}|$ with the spontaneously broken $U(1)$ symmetry of the overall phase and the superfluid order parameter $\chi_{+}^2$. {However, for the PP phase,}  we find a solution spontaneously chosen from a one-dimensional manifold of degenerate minima as illustrated in Fig.~\ref{fig:emf_j0.4}~(b).

In this phase of {two-superfluid components}, the PP phase, we may define another order parameter
\begin{align}
\chi_{-}^2 \equiv {\rm max} ||\Psi_{1}|^2-|\Psi_{2}|^2| .
\end{align}
For the case shown in Fig.~\ref{fig:emf_j0.4}, the maximum is realized for a vanishing relative phase $\Phi_{12}\to 0$.

In Figs.~\ref{fig:pd_p1.0}~(b) and (c), we show the full MF phase diagram of model~\eqref{BH2d}. Interestingly, there are several lobes of effective hardcore superfluid phases SF$_0$ and SF$_\pi$ separated by the intermediate PP phase which occupies a large part of the phase diagram for small $U/t$ and large fillings $\rho\gtrsim 1$. This phase diagram has to be compared to the DMRG results shown in Fig.~\ref{fig:pd_p1.0}~(a), which shows that in particular,
the extension of the MI phases is drastically underestimated by the MF-approximation.

With the solution Eq.\eqref{eq:MF_hardcore_psi}, we find \
\be
\chi_+^2 (n,n+1) = \frac{(zt)^2 (1+n)^2-(\mu-n U)^2}{2 (zt)^2 (1+n)}
\ee
and obviously $\chi_-^2$ and also the pair-density $\la b^2\ra$ vanish. {However, the PP phase} exhibits an enhanced pairing $\la b^2 \ra\sim \chi_-^2$. An example of the different order parameters for a cut through the phase diagram Fig.~\ref{fig:pd_p1.0} is shown in Fig.~\ref{fig:MF_cut}. We may interpret the PP phase as a phase of both a hardcore SF component and a partial formation of bound pairs on top of this background. Hence, this phase naturally extends the PP phase discussed for {three-body constrained anyons} in Ref.~\cite{Greschner2015}.
As already seen from the DMRG results, the PP phase has a {larger compressibility than the ordinary SF phases, however finite.}

Within the MF picture, the PP phase may also be defined away from $\theta=\pi$ by $\chi_+^2\neq 0$ and $\chi_-^2\neq 0$, while $\chi_-^2= 0$ for the SF$_0$ and SF$_\pi$ phases. Here, for $\theta\neq\pi$ the pairing $\la b^2 \ra$ is also non-vanishing for the SF phases, while $\chi_{-}^2=0$, which justifies the definition of the separate order parameter $\chi_{-}^2$.

\section{Effective repulsive interactions and stabilization for vanishing interactions} \label{sec:eff_int}

\begin{figure}[tb]
\centering
\includegraphics[width=\linewidth]{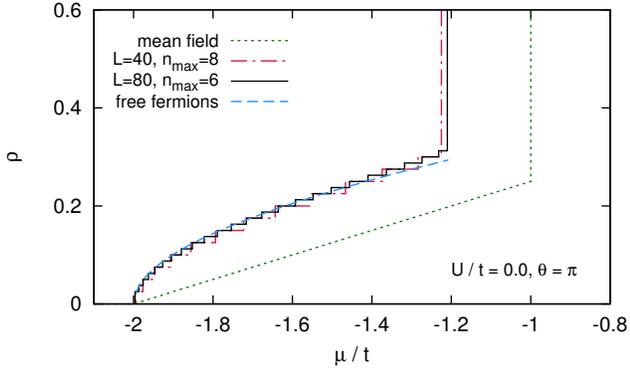}
\caption{Equation of state $\rho=\rho(\mu)$ for the AHM for  {a} statistical angle $\theta=\pi$ and vanishing on-site interactions $U=0$. Both DMRG and MF methods (setting $z=2$) show a stabilization of a SF phase for small fillings and a collapse after some critical filling factor. The MF case overestimated this point to $\mu/t=-1$, while the true value is around $\mu/t\approx-1.2$ as seen by the DMRG calculation.}
\label{fig:mag_U0_dmrg}
\end{figure}

The density dependent Peierls phase induces an effective repulsion which has important consequences on the {ground-state} phase diagram. A variation of the statistical angle $\theta$ may induce SF to MI transitions as observed in Refs.~\cite{Keilmann2011, Forero2016}.
In Refs.~\cite{Greschner2014, Greschner2015} this property has been explained from {the point of view} of a weak coupling analysis, in which the Luttinger liquid~(LL) parameter {is}
$K=\pi/(\theta^2+\frac{U}{2\rho t})^{1/2}$. Hence the statistical angle $\theta$ has qualitatively the same effect as a repulsive $U>0$.

We can explore this effect further in the MF-frame for the pseudo-fermion limit. For the low density case, the solution of \eqref{eq:MF_hardcore_psi}
for $\mu/t>-2$ is given by $\Psi \to \sqrt{4 t^2 - \mu^2}/(4 t)$ and the density is given
by $\la n\ra = (2+\mu/t)/4$ as shown in Fig.~\ref{fig:mag_U0_dmrg}.
For $\mu>-t$, however, the system becomes unstable and it is energetically favorable to form a macroscopically occupied site seen by the (infinitely) large jump in density in the $\mu$-$\rho$ curve.
Interestingly, the $\Psi_{0,1}$ solution remains stable also for small attractive interactions $U/t<0$.

So, remarkably, the fermionic off-site exchange statistics already induces a Pauli exclusion principle, i.e. a hardcore constraint for low fillings. We verify this property by means of DMRG calculations as shown in Fig.~\ref{fig:mag_U0_dmrg}. The low density part of the $\mu$-$\rho$ curve has {an} almost perfect overlap with the corresponding result from free (hardcore) fermions, for which $\rho={\rm arccos}(-\mu/2t)/\pi$.
The DMRG calculation indicates a slightly lower bound $\mu/t\approx -1.2$ for the instability of the system.

These findings are fully consistent with the analytic results known from anyonic continuum Lieb-Liniger models as presented e.g. in Refs.~\cite{Patu2007, Calabrese2007, Hao2008}. It has been shown that the original coupling constant $c$ is renormalized due to the exchange statistics $c'=c/\cos(\theta/2)$ resembling the described hardcore character for $\theta\to\pi$.
Also the two-particle scattering solution of the appendix~\ref{sec:app2particle} shows this effect - the two-particle scattering length Eq.~\eqref{eq:07_AHM_aAnyon} diverges for $\theta\to \pi$.

\section{Asymmetric momentum distribution} \label{sec:bsf}

\begin{figure}[tb]
\centering
\includegraphics[width=0.45\linewidth]{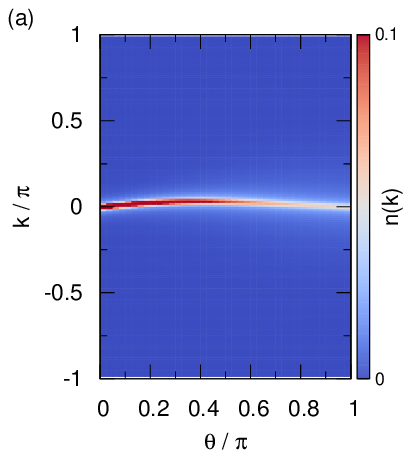}
\includegraphics[width=0.45\linewidth]{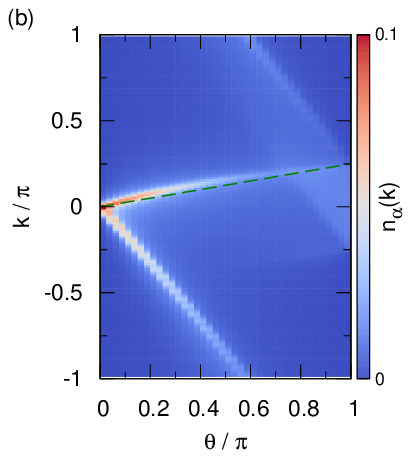}
\caption{Momentum distribution (a) of the bosonic particles $n(k)$ and (b) of the assumed anyons $n_\alpha(k)$ for the AHM (DMRG, $L=80$, $n_{max}=4$) for a small filling $\rho=1/4$ as a function of the statistical angle $\theta$.
The dashed line in panel (b) depicts the approximately linear dependence of the maximum position of Eq.~\eqref{eq:kmaxft}.}
\label{fig:dmrg_mom_b_a}
\end{figure}

\begin{figure}[tb]
\centering
\includegraphics[width=0.45\linewidth]{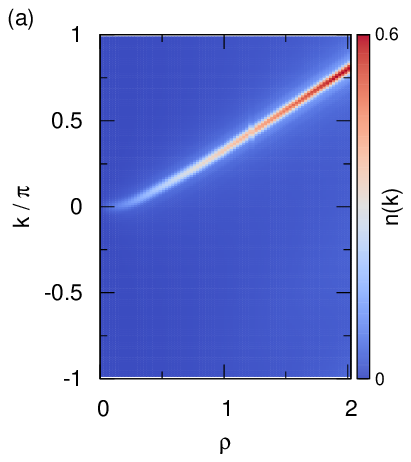}
\includegraphics[width=0.45\linewidth]{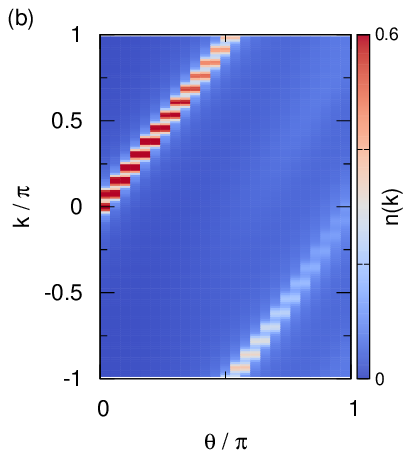}
\caption{(a) (Bosonic) momentum distribution $n(k)$ for the AHM (DMRG, $L=40$, $n_{max}=4$) as a function of the density for {a} statistical angle $\theta=\pi/2$.
(b) (Bosonic) momentum distribution $n(k)$ for the AHM (DMRG, $L=40$, $n_{max}=4$) as a function of the statistical angle $\theta/\pi$ for a density $\rho=93/40$.
The system is always in a superfluid state and we do not observe any phase {transitions} for these parameters.}
\label{fig:dmrg_asym_mom}
\end{figure}

A very important property of one-dimensional anyons discussed in Ref.~\cite{Keilmann2011}, is the asymmetric momentum distribution of the original bosonic particles, due to the broken space-reversal symmetry resulting from the phase factor assigned to the hopping.
As shown by Refs.~\cite{Hao2008,Keilmann2011,Tang2015}, the total bosonic momentum distribution $n(k)$
emerges in single peaks at some momentum $Q$ which are asymmetric with $k=0$. One may easily understand the asymmetry of the momentum distribution from the assumption of a system with a fixed density, $n_i \to \rho$. {Taking this into account}, the hopping-part of Hamiltonian~\eqref{BH} is given by
\be
\begin{aligned}
&\sum_i(b^{\dagger}_{i}b_{i+1}e^{i\theta \rho}+b^{\dagger}_{i}b_{i-1}e^{-i\theta \rho})\\
&=\sum_kb^{\dagger}_{k}b_{k}(e^{ik+i\rho\theta}+e^{-ik-i\rho\theta})
\end{aligned}
\ee
Therefore, $E(k)=-2t \cos (k+\theta \, \rho)$ and should be asymmetric with $k=0$ if $\theta \, \rho\ne 0$ .
One denotes this superfluid phase quasi condensing in the regime $0<Q<\pi$ due to the externally broken reflection symmetry
or  a broken symmetry superfluid (BSF) phase or generally SF$_Q$. In comparison to the PP phase which exhibits a complex momentum distribution function with several sharp peaks, there is only one distinct peak in the SF$_Q$ phases.

While here we are mainly interested in the experimentally observable properties of the underlying bosonic model, it is also an interesting question how the anyonic off-diagonal correlations and the corresponding momentum distribution,
\be
n_\alpha(k)=\frac{1}{L}\sum_{i,j}\langle \alpha_i^{\dagger}\alpha_{j}\rangle e^{ik(i-j)},
\label{nkalpha}
\ee
depend on the statistical angle $\theta$. It is important to note, that both quantities $n(k)$ and $n_\alpha(k)$ show completely different behaviors (contrary, for example, to the invariant density $n_i=\alpha_i^\dagger\alpha_i=b_i^\dagger b_i$). For example, for the case of impenetrable particles $U\to\infty$, the bosonic correlation functions are completely independent of the statistical angle. So $n(k)$ exhibits a single peak at $k=0$ while $n_\alpha(k)$ still shows a complex behavior interpolating between the (hardcore) boson and free fermion momentum distribution. These properties have been discussed extensively in Refs.~\cite{Santachiara2007, Hao2009, Calabrese2009, Tang2015, Marmorini2016}.

In Fig.~\ref{fig:dmrg_mom_b_a}, we compare both quantities $n(k)$ and $n_\alpha(k)$ for weakly interacting anyons ($U=0.5t$) for $\rho=1/4$. Again at sufficiently low fillings, we may compare the AHM to results obtained for continuum models~\cite{Calabrese2007, Patu2007}. The position of the single peak of the bosonic momentum distribution for these parameters is, to a first approximation, independent of the statistical angle. The anyonic momentum distribution, however, in Fig.~\ref{fig:dmrg_mom_b_a}~(b) shows a much more complex behavior. For small $\theta$, several peaks emerge. The position of the largest peak roughly follows
\be
k_{max}\sim \rho \theta
\label{eq:kmaxft}
\ee
in accordance with the analytical results of Ref.~\cite{Calabrese2007, Patu2007}.

In Figs.~\ref{fig:dmrg_asym_mom} {$\rm (a)$ and $\rm (b)$}, we present the momentum distribution of model~\eqref{BH} as {a} function of $\theta$ for $\rho\approx 2.3$ and as {a} function of the density $\rho$ for $\theta=\pi/2$ which illustrates its strong dependence on the filling and the statistical angle. In general,  SF$_0$, BSF and SF$_\pi$ phases are smoothly connected, and the peak position of the momentum distribution $Q$ continuously depends on density or statistical angle.

\begin{figure}[t]
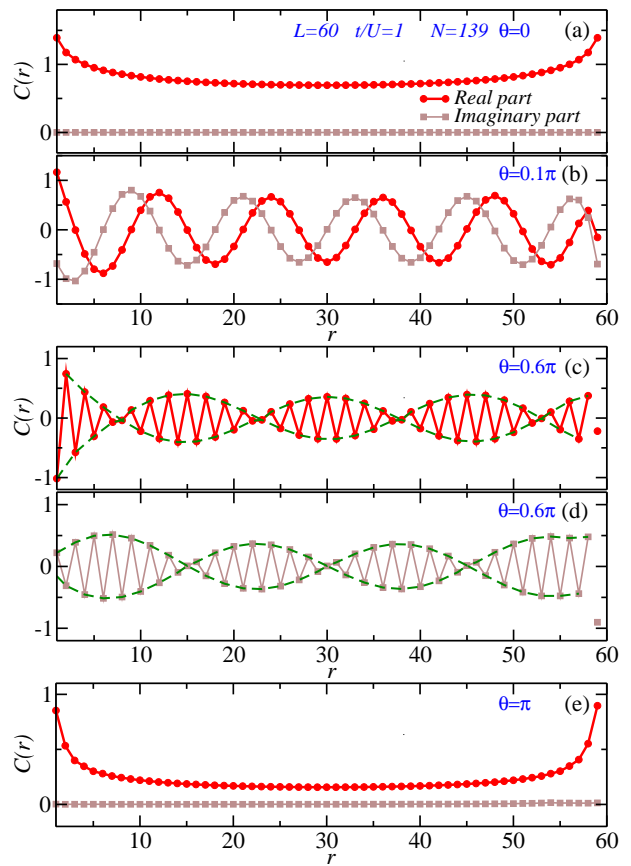

\includegraphics[width=0.45 \textwidth]{fig13ab.eps}
\includegraphics[width=0.45 \textwidth]{fig13cd.eps}
\includegraphics[width=0.45 \textwidth]{fig13e.eps}
\caption{Emergence and disappearance of beats from both the real and imaginary parts of the correlation $C(r)$ by modulation of $\theta$ for $\theta/\pi=0$, $0.1$, $0.6$ and $1$ at $t/U= 1$, $\rho=139/60$ DMRG data, $L=60$).}
\label{fig:c_rho2.3}
\end{figure}

In the following, we discuss properties of the single particle correlation function $C(r)=\langle b_i^{\dagger}b_{i+r}\rangle$ which exhibits an interesting {beat phenomenon} related to the asymmetry of the momentum distribution characteristic for the unconventional BSF phase.
For a SF$_Q$ superfluid phase at a quasi condensation at $Q$, with a momentum distribution peaked at $Q$ (see Fig.~\ref{fig:c_rho2.3}), we observe oscillatory patterns. For $\pi/2<Q\, {\rm mod}\, \pi<\pi$, both real and imaginary parts of the correlation function may exhibit a beat pattern as shown in Fig.~\ref{fig:c_rho2.3}~(c) and (d) as $\Re{C(r)} \sim \cos(Q r)$ and $\Im{C(r)} \sim \sin(Q r)$.
The correlations exhibit an algebraic decay typical for one-dimensional Luttinger liquids. Weak coupling properties of the (bosonic) single particle correlations have been derived in Ref.~\cite{Greschner2014}. The low-density properties of the anyonic correlations have been discussed in Refs.~\cite{Calabrese2007, Patu2007, Santachiara2007, Calabrese2009, Marmorini2016}.

\section{Conclusion}
\label{sec:con}

In summary, we have systematically studied the {ground-state} physics of the pseudo-AHM on a one-dimensional lattice without a constraint on the local particle number.
By using DMRG simulations, we have analyzed its characteristics and exotic properties. Some properties may be understood from a simplified
mean field analogy.

The main result of this paper is that close to the pseudo-fermion limit $\theta\approx\pi$, a partially paired phase of an atomic and a paired Luttinger-liquid component can be found to be stable for large fillings $\rho\gtrsim 1.5$. The simple mean field picture can be used for the illustration of this exotic quantum phase.

Furthermore, due to the broken space reflection symmetry, the momentum distribution gets asymmetrically shifted from $k=0$. We have shown that, while typically the momentum shifts smoothly with density, for $\theta=\pi$ there may be direct transitions between the SF$_0$ and SF$_\pi$ phases.

Finally, we have discussed how the anyonic exchange statistics  may induce an effective repulsion. For the case of pseudo-fermions and a statistical angle $\theta=\pi$, the fermionic off-site anti-commutation relations effectively generate a Pauli-exclusion principle.

The mean field analysis suggests that some of the features may be robust in more than {one dimension}. A two-dimensional variant of the pseudo-AHM Eq.~\eqref{BH2d}, even though its interpretation as an anyon-model would be inappropriate in this case, could be an interesting topic of further research.

\begin{acknowledgments}
We thank  Guixin Tang, Ming Gong, Luis Santos and Vladimir Korepin  for their invaluable discussions.
WZ and SG acknowledge the hospitality of KITPC Beijing.
WZ  is supported  by the NSFC under Grant No.11305113 and the Chinese Scholarship Council (CSC).
SG acknowledges support of the projects No. SA 1031/10-1 and No. RTG 1729 of the German Research Foundation (DFG).
{TCS} is supported in China by the project GDW201400042 for the ``high end foreign experts project''.
YZ is supported by NSF of China under Grant Nos. 11674201 and 11474189.
Simulations were partially carried out on the cluster system at the Leibniz University of Hannover, Germany.
\end{acknowledgments}

\appendix

\section{Two-particle analysis of the AHM}\label{sec:app2particle}

As already shown in Refs.~\cite{Greschner2015, Cardarelli2016} in the dilute limit $\rho\to 0$, we may derive a description of the properties of the AHM by means of a two-particle scattering problem~\cite{Kolezhuk2012}.
A general two-particle state may be described by
\begin{equation}
\left| \Psi_K \right> = \sum_x c_{x,x} \left(b_x^\dagger\right)^2 \left|0\right> + \sum_{x,y>x} c_{x,y} b_x^\dagger b_y^\dagger \left|0\right>.
\end{equation}
We may express the amplitudes as $c_{x,x+r} = C_r \e^{i Q (x+\frac{r}{2})}$ due to the conservation of total momentum  $Q=k_1+k_2$ in the scattering process.
{Taking this into account}, the Schr\"odinger equation $H \left|\Psi\right> = \Omega \left|\Psi\right>$ reduces to the following system of coupled equations:
\begin{widetext}
\begin{eqnarray}
&&\!\!\!\!\!\!\!\!\!\!(\epsilon_2 - U) C_0 = - \sqrt{2} t \left( \e^{-\ii \frac{Q}{2}} + \e^{\ii \left(\frac{Q}{2} + \theta\right)} \right) C_1 \label{eq:dilute_2particle_eqs-1}\\
&&\!\!\!\!\!\!\!\!\!\!\epsilon_2 C_1 = - \sqrt{2}  t \left( \e^{\ii \frac{Q}{2}} + \e^{-\ii \left(\frac{Q}{2} + \theta\right)} \right) C_0 + 2 t \cos\left(\frac{Q}{2}\right) C_2 \label{eq:dilute_2particle_eqs-2} \\
&&\!\!\!\!\!\!\!\!\!\!\epsilon_2 C_r = -2  t \cos\left(\frac{Q}{2}\right) \left( C_{r-1} + C_{r+1}\right)\, ,r\geq 2 \label{eq:dilute_2particle_eqs-3}
\end{eqnarray}
\end{widetext}

Let us first consider scattering states of two particles for which, in the thermodynamic limit, the energy is given by
$\Omega = \epsilon(k_1) + \epsilon(k_2) = - 4 t \cos(q) \cos\left(\frac{Q}{2}\right)$
where $q=(k_1-k_2)/2$. We may solve this set of equations with the ansatz $C_r = e^{-\mathrm{i} q r} + e^{2\mathrm{i}\delta} e^{\mathrm{i} q r}$.
The coefficients $C_0$ and $\delta$ are determined by Eqs.~\eqref{eq:dilute_2particle_eqs-1} and \eqref{eq:dilute_2particle_eqs-2}
and hence, are affected by the interactions and anyon statistics.
From the scattering phase shift $\delta$, we can extract the scattering length,
\begin{equation}
a = \frac{t (1+\cos\theta)}{-2 (2 t+U)+4 t \cos\theta} \;.
\label{eq:07_AHM_aAnyon}
\end{equation}
By comparison to a 1D Bose gas of particles with mass $m$ and contact interaction, one may identify $a$ with an effective interaction strength $g=-2/(a m)$ ~\cite{Kolezhuk2012}.
The scattering length diverges for $\theta\to0, 2\pi$ but remains finite and negative for any other phase $\theta$. This again shows the effective repulsion induced by the anyonic exchange statistics. For $\theta\to\pi$, the scattering length $a\to 0$ and, hence, the system approaches the Tonks limit $K\to 1$ of a hardcore free (fermion) gas as already discussed in section~\ref{sec:eff_int}.

\begin{figure}[tb]
\centering
\includegraphics[width=\linewidth]{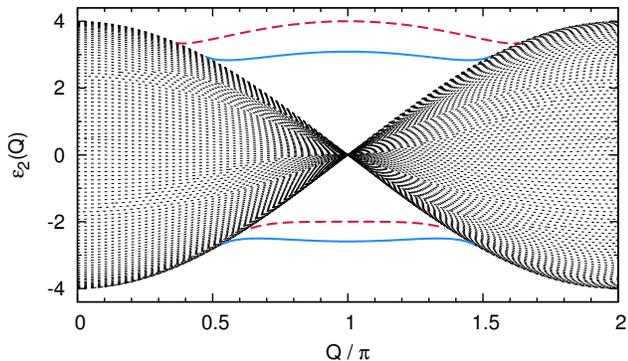}
\caption{Emergence of bound states in the two-particle spectrum $\epsilon_2(k)$ of the AHM for $\theta=\pi$. Dotted black lines illustrate the energies of two-particle scattering solutions. The solid (blue) and dashed (red) lines denote the stable bound-state solutions for repulsive interactions $U=0.5 t$ and $U=2t$ resp.}
\label{fig:dilute_scattering}
\end{figure}

An apparently counterintuitive observation is, that in spite of this effective hardcore character of the two particle scattering state, {nonetheless} low-lying bound states of two particles may exist (see also discussion in Ref.~\cite{Cardarelli2016}). In order to analyze the two-particle {bound states}, we use the ansatz $C_r=\alpha^r$ with $|\alpha|<1$ such that the solution is exponentially localized to its center of mass. We may again solve Eqs.~\eqref{eq:dilute_2particle_eqs-1}-\eqref{eq:dilute_2particle_eqs-3} with this ansatz. Note, that for the usual Bose-Hubbard model, $\theta=0$ for repulsive interactions $U>0$, stable solutions of repulsively bound-pairs is found only for high energies above the two-particle scattering spectrum.
However, the AHM for $\theta\to\pi$ for any $U$ (repulsive or attractive) exhibits two different {bound-state} solutions close to $Q\sim\pi$ as shown in Fig.~\ref{fig:dilute_scattering}, with energies given by
\begin{align}
\epsilon_\pm^B = \frac{2U\cos(k)\pm\left(\cos(k)-1\right)\sqrt{U^2 + 8 t^2 (1 - 3 \cos(k))}}{3 \cos(k)-1}
\end{align}
Interestingly, for any $U>0$, one of these solutions
has energies inside the two-particle spectrum $\epsilon_+^B < 0$. For $U<2t$, it exhibits a local minimum at $Q=\pi$.

As discussed in Refs.~\cite{Greschner2015,Cardarelli2016}, we may now understand the exotic PP phase in a simplified picture as a phase of the simultaneous presence of both a gas of strongly repulsively interacting unpaired particles and a quasi-condensate of pairs (corresponding to the minimum of the low-lying bound state solution). Since, $\epsilon_+^B$ is not the lowest energy of the {two-particle} solution (as long as $U>-2t$), we may not expect a pure pair (quasi)-condensate such as the PSF phase studied for attractive interactions~\cite{Daley2009}.
Neglecting interactions between these two quasi-particles, which may be reasonable for small densities of atoms $\rho_a$ and pairs $\rho_p$, one may write a simplified model of the AHM as
\begin{align}
H_{eff} = - 2 t \sum_k \cos(k) a_k^\dagger a_{k} + \sum_k \epsilon_-^B(k) b_k^\dagger b_{k}
\label{eq:07_AHM_ham_ad}
\end{align}
where $a_k$ ($a_k^\dagger$) and $b_k$ ($b_k^\dagger$) are annihilation (creation) operators of hardcore atom and bound-pairs of momentum $k$.
The Hamiltonian has to be minimized under the constraint $\rho_a + 2\rho_p = \rho$. Although being certainly an oversimplification, Model~\eqref{eq:07_AHM_ham_ad} captures some main physical aspects: at low densities the ground state only contains species $a$; for higher fillings both species are present.

We may expand this dilute limit analysis to larger fillings by assuming the presence of a uniform and constant background filling $n$. Hence, the single particles moving on top of this background $|n+1\rangle$ and the bound doublon pairs $|n+2\rangle$ obtain a renormalized hopping rate due to the bosonic enhancement. {By repeating the above analysis} of bound and scattering states, we obtain a qualitatively similar picture for small $n=1,2,\cdots$.


\end{document}